\documentclass[preprint,pteplogo]{ptephy_v2}
\usepackage{url}

\preprintnumber{XXXX-XXXX} 
\usepackage{hyperref}
\usepackage{graphicx}
\usepackage{amsmath}
\usepackage{subcaption}
\usepackage{float}
\usepackage{placeins}  
\usepackage{afterpage}
\usepackage{newunicodechar}
\usepackage{CJK}

\begin{document}

\title{A Compact Two-Dimensional Radiation Detector for Educational Applications}


\author[1,2]{Chiori Matsushita}
\author[1,3]{Mihiro Nukiwa}
\author[1,4]{Aoi Atobe}
\author[1,5]{Yuzuka Sasaki}
\author[1,6]{Manami Sawai}
\author[1,7]{Rikako Kono\thanks{Corresponding author rikako.kono@anu.edu.au}}
\author[8]{Seyma Esen}
\author[9]{Martin Schwinzerl}
\author[1,10]{Kazuo S. Tanaka}

\affil[1]{Accel Kitchen LLC \email{info@accel-kitchen.com}}
\affil[2]{Joshigakuin Junior and Senior High School}
\affil[3]{Institute of Science Tokyo}
\affil[4]{Junten Senior High School}
\affil[5]{UWC Adriatic}
\affil[6]{Aeronautical Safety College}
\affil[7]{Australian National University}
\affil[8]{CERN, Istanbul University}
\affil[9]{CERN, German Center for Astrophysics}
\affil[10]{Waseda University}




\begin{abstract}%
Recent advances in low-cost, portable cosmic-ray detectors have broadened citizen-science engagement in particle physics studies. Imaging applications such as muography, however, remain largely inaccessible because existing detectors typically require tens to hundreds of sensors, making them costly and complex. This gap underscores the need for a compact imaging detector.

In this study, we developed SAKURA, a palm-sized two-dimensional muon scintillation detector that uses only four silicon photomultipliers (SiPMs) and can be built for less than 1,000 USD. Its spatial resolution was evaluated using a 5 GeV/c muon beam at CERN's T10 beamline in September 2024, yielding 13.4 mm along the x-axis and 7.48 mm along the y-axis. The entire study, including designing and testing the detector, and analysing data, was undertaken by high school students, demonstrating that SAKURA makes radiation imaging accessible and practical even for non-specialists with the help of some professional scientists.
\end{abstract}

\subjectindex{Muography, Radiation imaging, Scintillation detector, Citizen science}

\maketitle

\section{Introduction}\label{sec:intro}
    Compact particle detectors that pair silicon photomultipliers (SiPMs) with scintillators have recently gained increasing attention among citizen scientists. Devices such as Cosmic Watch \cite{CosmicWatch} and Radiacode \cite{Radiacode} make particle-physics research accessible to people without access to dedicated laboratory setups through their portable and low-cost designs \cite{appliedPhys_sakura,appliedPhys_tanaka}. These detectors enable conventional radiation studies, such as the east-west effect \cite{PhysRev.75.590} and the zenith angle distribution of cosmic rays \cite{PhysRev.93.590}, but they cannot track particles because they detect only the integrated output of a monolithic scintillator, limiting the scope of citizen science projects. \par

    Particle tracking enables such applications as muography, a form of tomography that exploits variations in muon attenuation through matter. Muography is a well-established technique and has uncovered hidden chambers inside the Great Pyramid of Giza \cite{Morishima2017,Procureur2023} and mapped the interiors of volcanoes for the forecast of volcanic eruptions \cite{Tanaka2019,LoPresti2020}. Existing muography systems, however, rely on the order of 50 to 150 sensors \cite{Tanaka2010,ANGHEL201512,Saracino2017}, making them expensive, cumbersome, and complex to operate; their use is confined to a handful of research institutions. \par
    
    For the application of muography to inspect small-scale infrastructures, such as water pipes or tunnels, several groups have developed portable muon trackers \cite{PhysRevResearch.2.023017,10.1063/1.4922006,Guardincerri2017}. Existing trackers, however, span roughly 1 m or more and remain insufficiently portable for many field applications, underscoring the need for more compact imaging solutions. \par
    
    To address these challenges, we have developed a low-cost, portable, and low-power two-dimensional imaging detector named SAKURA, derived from Cosmic Watch~\cite{CosmicWatch}. The detector consists of a $5\times5$ array of caesium iodide (CsI) scintillators and four silicon photomultipliers (SiPMs) positioned on the top, bottom, left, and right sides of the array. The position of the scintillation event is reconstructed by analysing the relative signal strengths from these four SiPMs. Measuring approximately 50~mm and operating on less than 1~W via a USB power source, SAKURA can be built for under 1{,}000~USD.
    
    The detector was selected as one of the winning proposals in the 2023 edition of the international high school competition \href{https://beamlineforschools.cern/}{\textit{Beamline for Schools (BL4S)}}, organised by CERN in collaboration with DESY, and was tested at the T10 beamline in muon configuration in CERN's East Area in September 2024~\cite{cern_bl4s_winners_2024,appliedPhys_sakura}. The entire process---from development and construction to beam testing and data analysis---was carried out by high school students, making SAKURA highly suitable for educational purposes as a detector that students can build and operate on their own.

\section{SAKURA detector}\label{sec:detector}
\begin{figure}[t]
        \centering
        \includegraphics[scale=0.65]{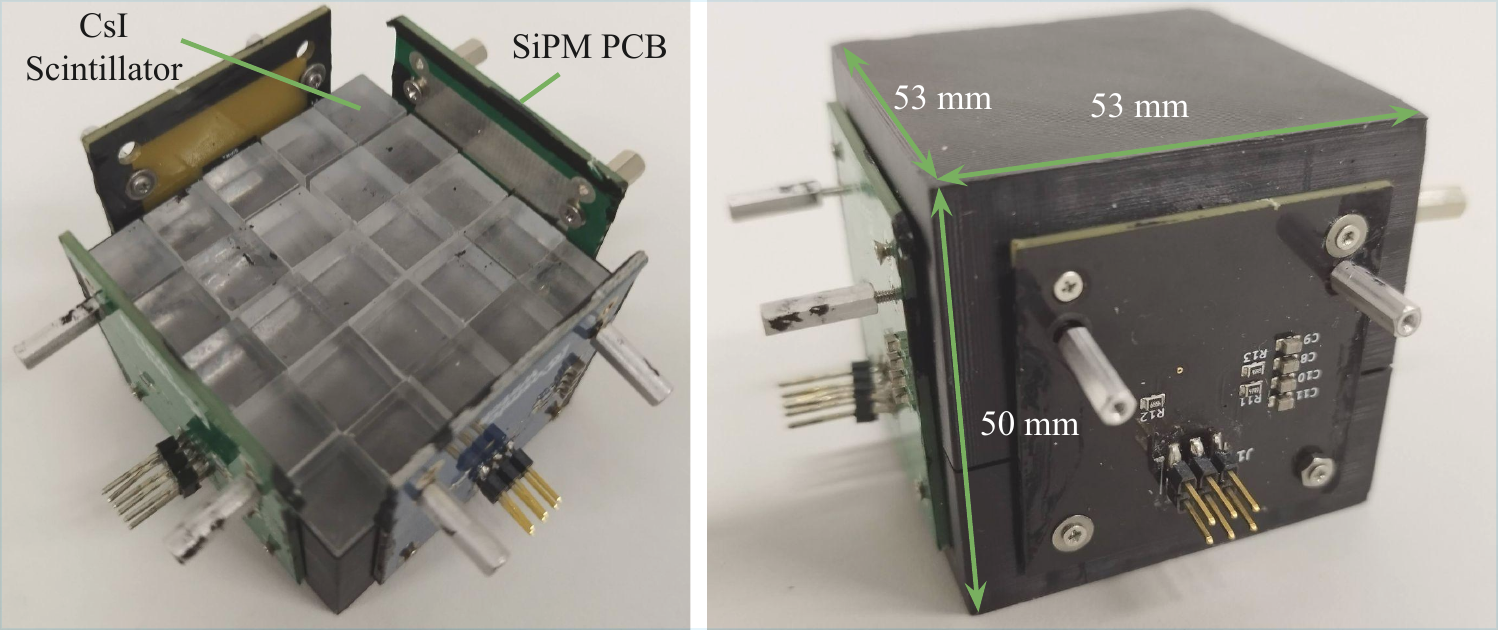}
        \caption{
                    SAKURA consists of twenty-five $10 \times10\times10~\text{mm}^3$ CsI scintillators arranged in a $5 \times 5$ grid, with four SiPMs mounted on its four lateral faces. Scintillation light emitted from the irradiated scintillator propagates through the air gaps, with its intensity attenuated, causing the signal intensity recorded by each SiPM to depend on the incident position.}
        \label{fig:SAKURA_scheme}
    \end{figure}

    SAKURA is a two-dimensional position-sensitive detector developed based on the desktop muon detector Cosmic Watch~\cite{CosmicWatch}. While Cosmic Watch employs a single $50\times50\times10~\text{mm}^3$ plastic scintillator read out by one SiPM, SAKURA replaces the monolithic crystal with a $5\times5$ array of twenty-five $10\times10\times10~\text{mm}^3$ CsI scintillators, as shown in Fig.~\ref{fig:SAKURA_scheme}. Four Onsemi SiPMs (MICROFC-60035-SMT-TR1) are mounted on the four lateral faces of the array, one on each side.

    All components are housed in a light-tight inner case measuring $53\times53\times50~\text{mm}^3$, which is enclosed within an outer casing with a side length of $184~\text{mm}$. A 5~V power supply from USB is boosted to 29.5~V and applied as the bias voltage to the SiPMs. The SiPM output signals are shaped to a pulse width of approximately 0.5~$\mu$s using an RC filter, and then converted into voltage through load resistors for readout. The voltage signals are acquired using Red Pitaya (STEMlab 125-14 4-Input), an ADC+FPGA board.
    
    Red Pitaya is a compact, USB-powered board widely used for educational purposes. It allows data acquisition scripts to be implemented directly via a web console using Jupyter Notebook. This setup makes it easy even for high school students to write their own analysis code, with previous successful applications including the characterisation of Cherenkov detectors by students~\cite{jofscieggs-cherenkov}. Thanks to its 125~MHz ADC sampling, Red Pitaya enables the full waveforms of all four channels to be recorded for each event, allowing real-time reconstruction based on waveform data. During the evaluation at CERN using a muon beam, an oscilloscope was also employed in parallel with the Red Pitaya for data acquisition, as described in Sec.~\ref{sec:method}.

    Position reconstruction is based on the following principle:
        When radiation passes through SAKURA, the scintillator emits light that propagates into adjacent crystals before being detected by the four SiPMs. Because this light is progressively attenuated through the air gaps between scintillators, the signal intensity recorded by each SiPM depends on the position of incidence. A straightforward centre-of-gravity (CoG) method therefore reconstructs the coordinates. 
        \begin{figure}[t]
            \centering
            \includegraphics[scale=0.7]{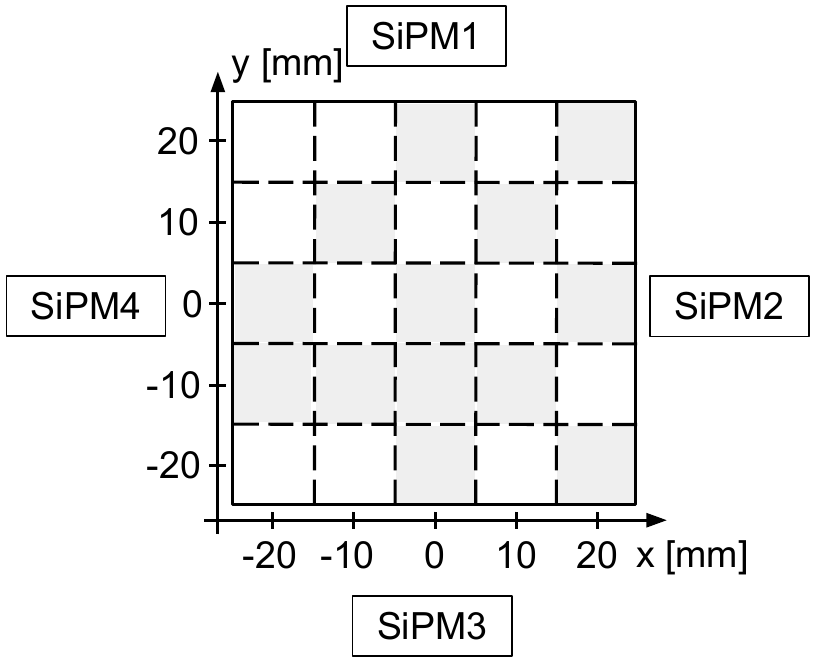}
            \caption{
            Front view diagram of SAKURA. The coordinate origin is defined at the centre of the scintillator array. SiPM1 is located at the top, followed clockwise by SiPM2, SiPM3, and SiPM4. The thirteen grey squares indicate the predefined positions where the muon beams were directed.}
            \label{fig:ch_assignment}
        \end{figure}
        Let $V_{i}$ be the peak amplitude from SiPM$_i$ (see Fig. \ref{fig:ch_assignment} for channel labels) and ($x_i$, $y_i$) its coordinates. The incident position (X, Y) is then calculated as follows:
        \begin{align}
            X\;[\text{mm}] &= \frac{\sum_{i=1}^{4} x_i V_i}{\sum_{i=1}^{4} V_i}
                 = \frac{25\,(V_2 - V_4)}{V_1+V_2+V_3+V_4}, \label{eq:x_loc_calc}\\[4pt]
            Y\;[\text{mm}] &= \frac{\sum_{i=1}^{4} y_i V_i}{\sum_{i=1}^{4} V_i}
                 = \frac{25\,(V_1 - V_3)}{V_1+V_2+V_3+V_4}. \label{eq:y_loc_calc}
        \end{align}

\section{Detector Performance Evaluation}\label{sec:method}

\begin{figure}[t]
        \centering
        \includegraphics[scale=0.65]{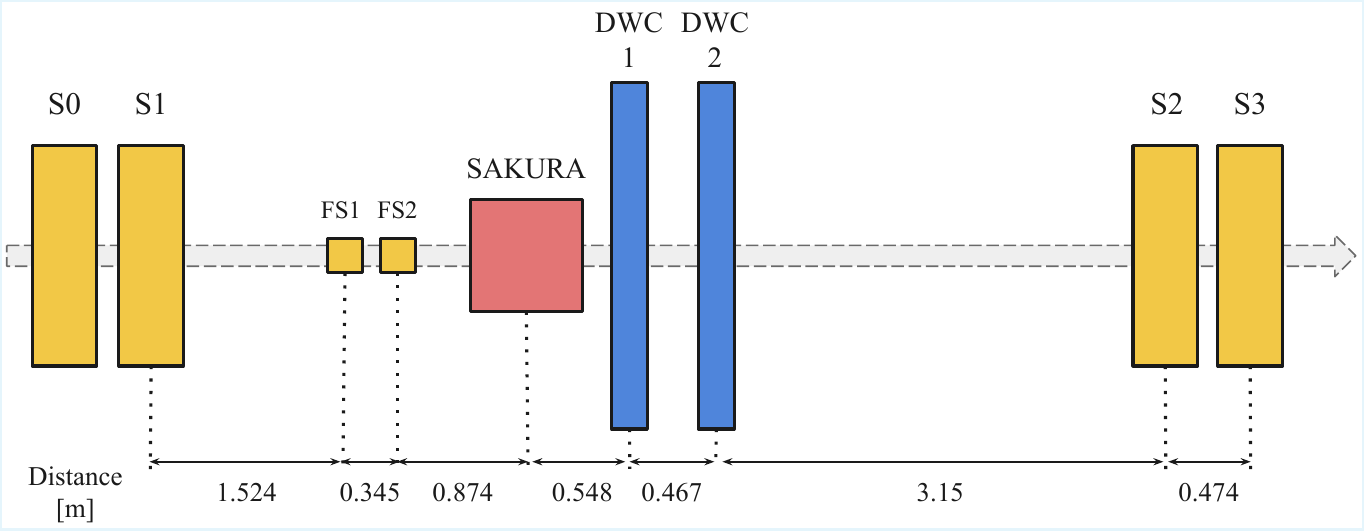}
        \caption{
                Diagram of the experimental setup at the CERN T10 beamline. Two DWCs were positioned downstream of SAKURA for reference measurement.
                Two trigger systems were used: one using plastic scintillation detectors (S0–S3) for data acquisition, and the other using finger scintillators (FS1 and FS2) for offline analysis.}
        \label{fig:beamline_scheme}

    \end{figure}

    An experiment was conducted at CERN's T10 Proton Synchrotron (PS) test beamline in September 2024 to evaluate the spatial resolution of SAKURA. A 5 GeV/c muon beam delivered $\approx2\times10^3$ events per 0.4-second spill. Two $100\;\text{mm}\times100\;\text{mm}$ Delay Wire Chambers (DWCs) with a spatial resolution of 0.25 mm were installed downstream of SAKURA as illustrated in Fig.~\ref{fig:beamline_scheme}; their measurements served as a benchmark for the muon positions in the SAKURA plane. SAKURA waveforms were recorded with a Tektronix MSO58LP oscilloscope and fed into the data acquisition system shared with DWCs, enabling event-by-event comparison between SAKURA and DWCs. Online triggers were provided by plastic scintillators (S0–S3), while a two-fold coincidence of finger scintillation detectors, FS1 and FS2, was applied during offline analysis.

    The muon beam was directed to thirteen predefined positions on SAKURA, as indicated by the grey squares in Fig.~\ref{fig:ch_assignment}.
        \par
\begin{figure}[t]
        \centering
        \includegraphics[scale=0.6]{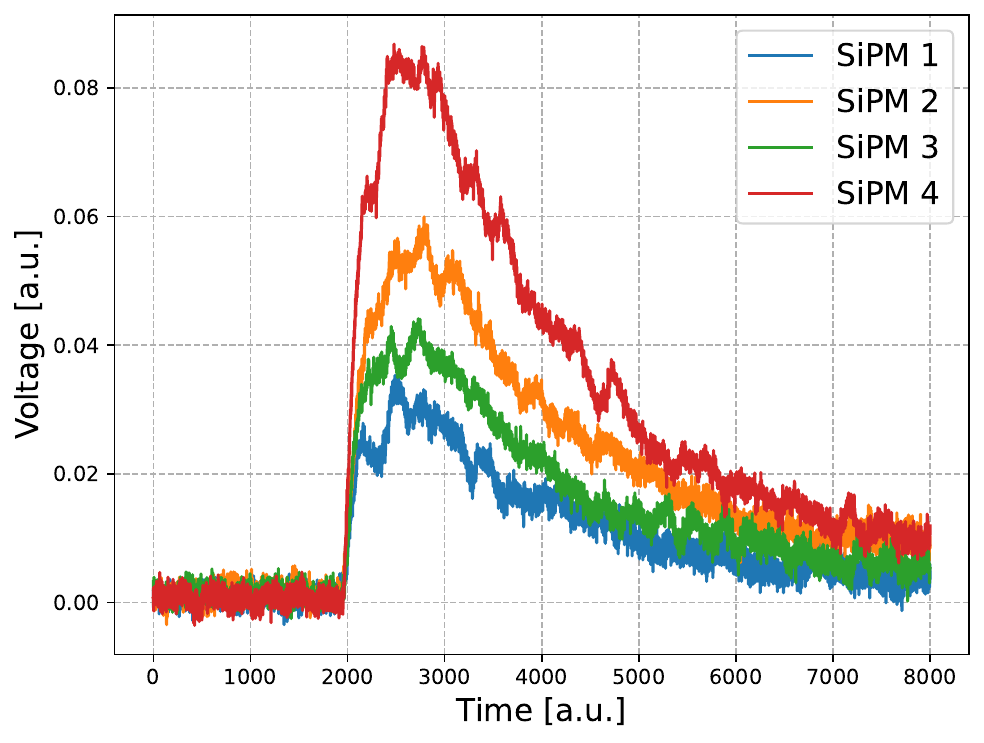}
        \caption{
                Representative oscilloscope waveforms from the four SiPMs. Events showing double peaks or saturation were excluded from the subsequent analysis. This example suggests that the muon hit the scintillator near to SiPM4.}
        \label{fig:oscilloscope_waveform}
    \end{figure}
\begin{figure}[t]
        \centering
        \includegraphics[scale=0.53]{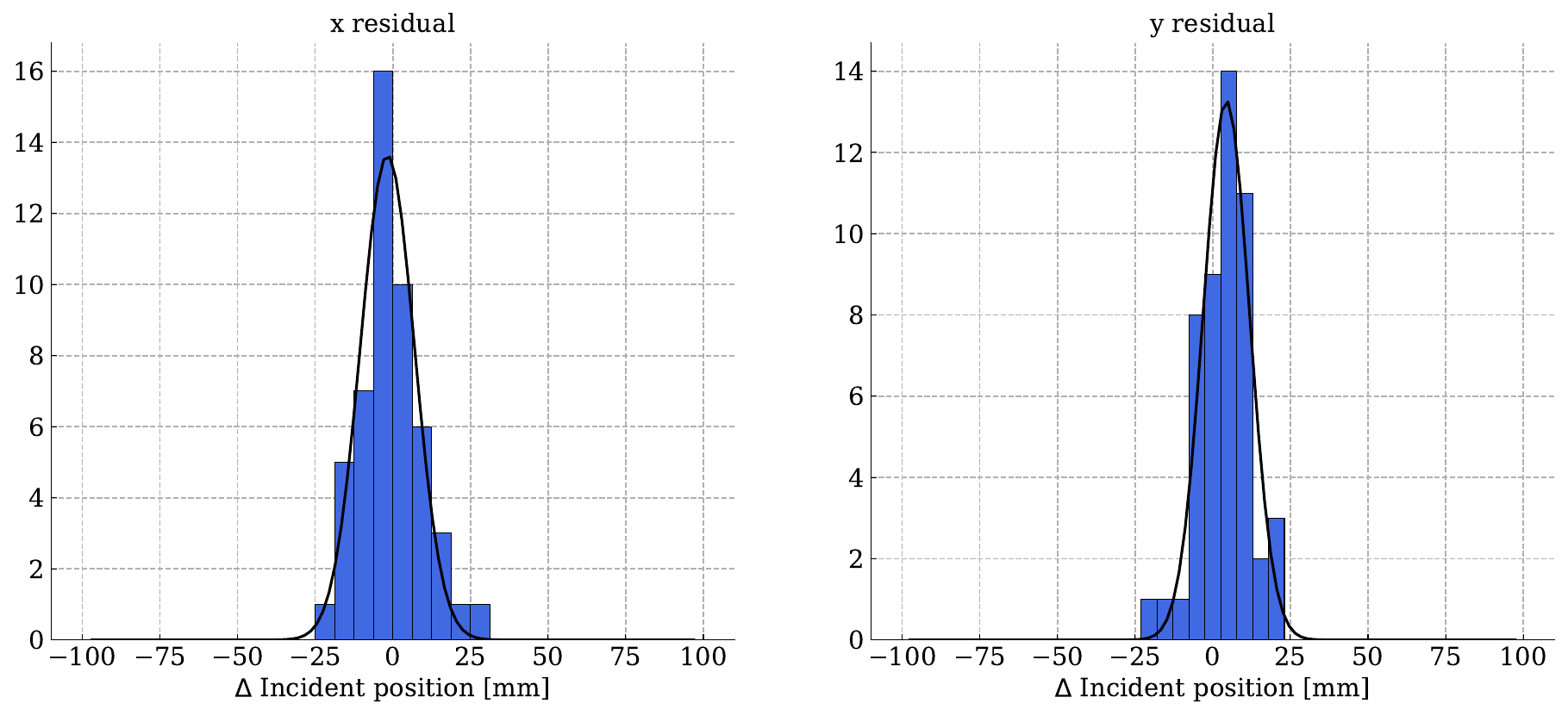}
        \caption{                   
                    The histograms of the residual between SAKURA and DWC-estimated positions at (X, Y) = (10 mm, -10 mm). Each histogram was fitted with a Gaussian using a least-squares method.}
        \label{fig:residual}
    \end{figure}
\begin{figure*}[t]
        \centering
        \begin{subfigure}[b]{0.49\textwidth}
            \centering
            \includegraphics[width=\textwidth]{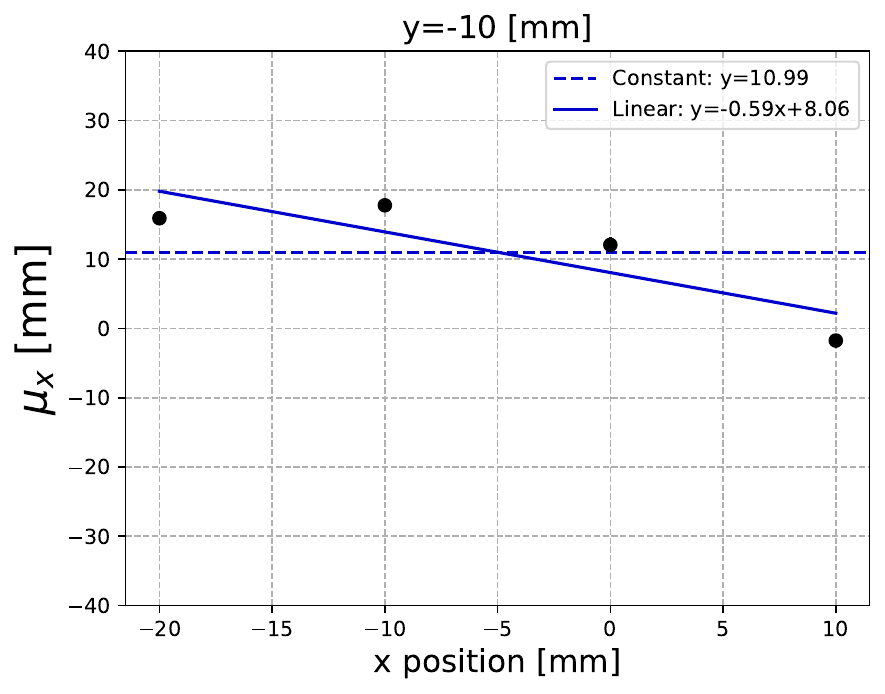}
        \end{subfigure}
        \hfill
        \begin{subfigure}[b]{0.49\textwidth}  
            \centering 
            \includegraphics[width=\textwidth]{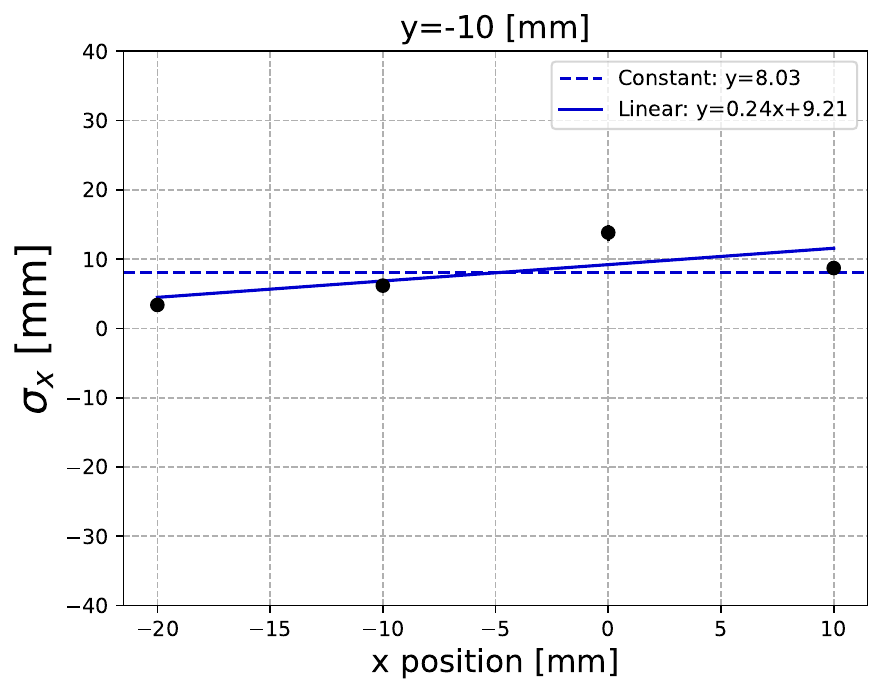}
        \end{subfigure}
        \vskip\baselineskip
        \begin{subfigure}[b]{0.49\textwidth}   
            \centering 
            \includegraphics[width=\textwidth]{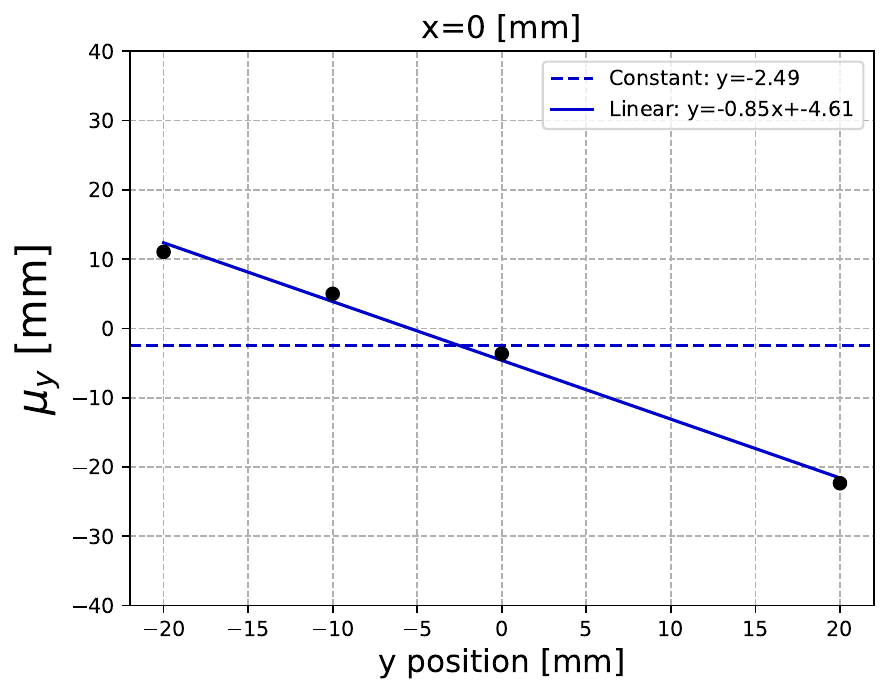}
        \end{subfigure}
        \hfill
        \begin{subfigure}[b]{0.49\textwidth}   
            \centering 
            \includegraphics[width=\textwidth]{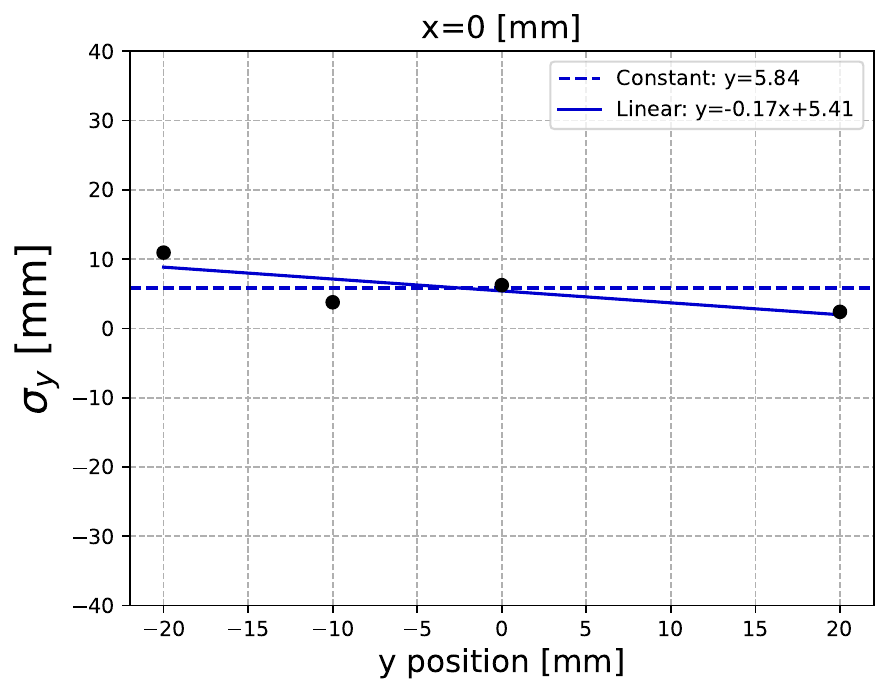}
        \end{subfigure}
        \caption{
            Measured $\mu$ and $\sigma$ values from residual histograms on the $x=0\;\text{mm}$ and $y=-10\;\text{mm}$ axes. The data are fitted with a constant model (dotted line) and a linear model (solid line) to evaluate position dependence.}
        \label{fig:sigma_mu}
        \vspace{-10pt}
    \end{figure*}

    To determine the spatial resolution of SAKURA and its dependence on position, we compared muon positions reconstructed by SAKURA with those measured by the DWCs. Fig.~\ref{fig:oscilloscope_waveform} displays representative oscilloscope waveforms from SiPM1 to SiPM4. Waveforms exhibiting double peaks or saturation were excluded from the analysis. 
    Fig.~\ref{fig:residual} shows the residual distributions between SAKURA and the DWC-estimated positions at \( (X, Y) = (10 \,\mathrm{mm},\ -10 \,\mathrm{mm}) \). Each distribution was fitted with a Gaussian function using the least-squares method, and the same analysis was repeated for all thirteen beam positions.
    The resulting mean ($\mu$) and standard deviation ($\sigma$) at positions along the vertical line at $x=0\;\text{mm}$ and the horizontal line at $y=-10\;\text{mm}$ are shown in Fig.~\ref{fig:sigma_mu}. To investigate the position dependence of these values, we fitted both a constant (position-independent) model and a linear (position-dependent) model to the data and compared them using an F-test under the null hypothesis of no position dependence. The F-statistic was calculated from the sum of squared residuals (SSR) and degrees of freedom (df) for the two models, as defined in Eq.~\ref{eq:f_test}. The fitted parameters, F-statistics, and corresponding p-values are summarised in Tab.~\ref{tab:f_test_result}.
    
    \begin{align}
        \text{F}=\frac{\text{SSR}_{\text{const}}-\text{SSR}_{\text{linear}}}{\text{df}_{\text{const}}-\text{df}_{\text{linear}}}/\frac{\text{SSR}_{\text{linear}}}{\text{df}_{\text{linear}}}
        \label{eq:f_test}
    \end{align}
    
    \begin{table}[h]
        \centering
        \caption{
                Comparison of constant and first-order fits to the mean ($\mu$) and standard deviation ($\sigma$) measured along the x=0 mm vertical and y=$-10$ mm horizontal axes. For each parameter, the table lists the best-fit coefficients, F-statistic, and p-value used to test the null hypothesis of no position-dependent trend (significance threshold p=0.05).}
    \begin{tabular}{ccccc}
         & Constant fit & Linear fit & F value & p value \\\hline\hline
        $\sigma_x$ & $\sigma_x = 8.03~\mathrm{mm}$ & $\sigma_x = 0.236\,x + 9.21~\mathrm{mm}$  & 1.79 & 0.313 \\
        $\mu_x$    & $\mu_x = 11.0~\mathrm{mm}$    & $\mu_x = -0.587\,x + 8.06~\mathrm{mm}$    & 5.60 & 0.142 \\
        $\sigma_y$ & $\sigma_y = 5.84~\mathrm{mm}$ & $\sigma_y = -0.172\,y + 5.41~\mathrm{mm}$ & 3.12 & 0.219 \\
        $\mu_y$    & $\mu_y = -2.49~\mathrm{mm}$   & $\mu_y = -0.848\,y - 4.61~\mathrm{mm}$    & 281  & 0.00354 \\
    \end{tabular}
        \label{tab:f_test_result}
    \end{table}

    Because the p-values for $\mu_x$, $\sigma_x$, and $\sigma_y$ all exceeded 0.05, the null hypothesis could not be rejected, indicating no significant dependence on position. By contrast, the p-value for $\mu_y$ was 0.00354; the null hypothesis was therefore rejected, indicating a significant position-dependent trend. This systematic shift is plausibly attributed to a slight downward tilt of the scintillator due to the weight of  SAKURA itself. Based on these results, we corrected the residual distributions along the x and y axes using the following equations:

    \begin{align}
        \text{Residual}_{x,\text{modified}}=\text{Residual}_{x,\text{original}}-11.0.
        \label{eq:adjustment_x}
    \end{align}
    \begin{align}
        \text{Residual}_{y,\text{modified}}=\text{Residual}_{y,\text{original}}-(-0.848y-4.61).
        \label{eq:adjustment_y}
    \end{align}

\begin{figure}[t]
        \centering
        \includegraphics[scale=0.53]{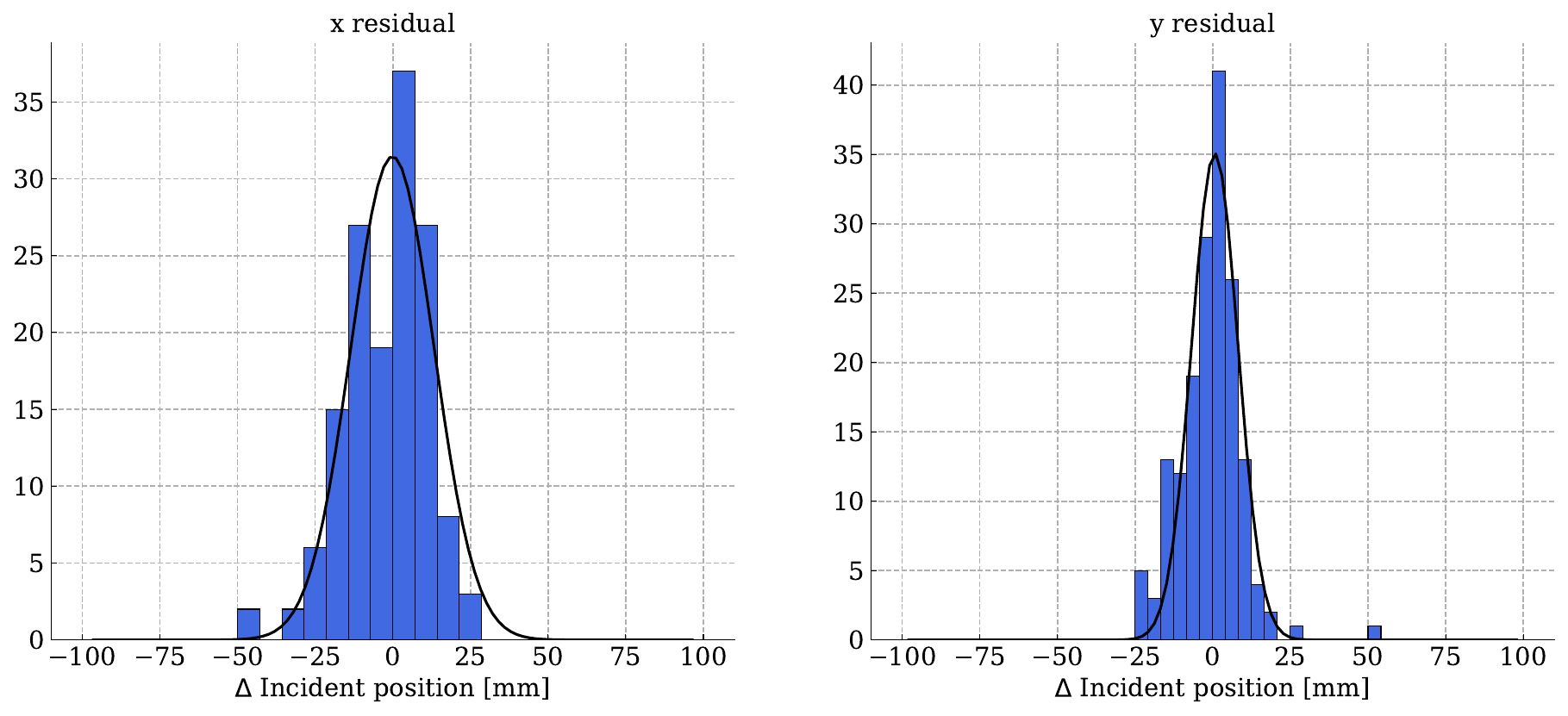}
        \caption{
                    Combined residual histograms along $x=0\;\text{mm}$ and $y=-10\;\text{mm}$ lines after position correction. Gaussian fits are applied to the histograms, yielding $\sigma_x$ = 13.4 mm and $\sigma_y$ = 7.48 mm.}
        \label{fig:residual_combine}
    \end{figure}

    Since no systematic dependence was observed in the $\sigma$ values, the data at incident positions along the lines $x=0\;\text{mm}$ and $y=-10\;\text{mm}$ were combined, as shown in Fig.~\ref{fig:residual_combine}. Gaussian fits were then applied to the combined distributions, yielding $\sigma_x$ = 13.4 mm and $\sigma_y$ = 7.48 mm. 
    Using these corrected $\sigma$ values, the spatial resolution of SAKURA along each axis was calculated using

    \begin{align}
        \sigma_{\text{SAKURA}}=\sqrt{\sigma_{\text{residual}}^2-\sigma_{\text{DWC}}^2}.
        \label{eq:combined_sigma}
    \end{align}

\begin{figure}[t]
        \centering
        \includegraphics[scale=0.55]{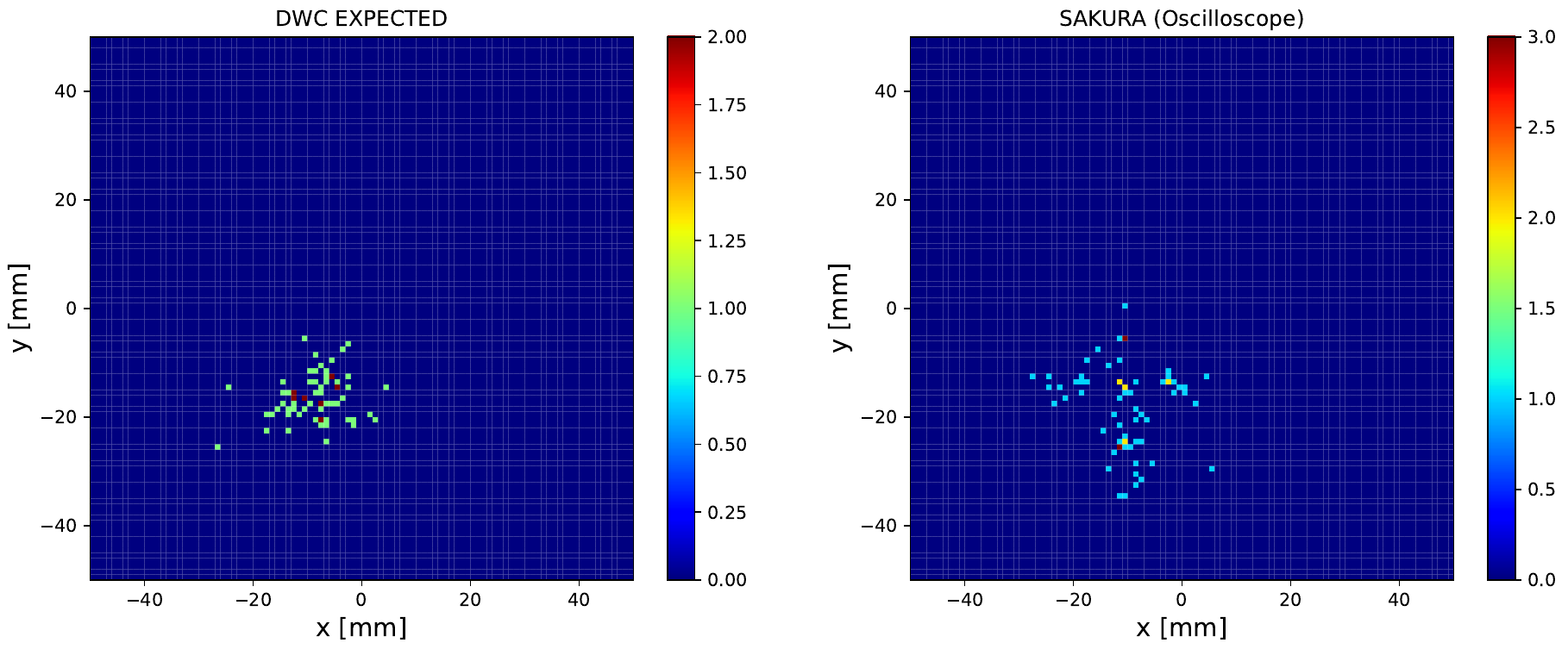}
        \caption{
                  Corrected hitmaps at (X, Y) = (0 mm, -20 mm) from DWC and SAKURA, showing improved agreement with the DWC reference coordinates.}
        \label{fig:corrected_hitmap}
    \end{figure}

    Given the spatial resolution of DWC of 250 $\mu m$, the spatial resolution of SAKURA was determined to be 13.4 mm along the x-axis and 7.48 mm along the y-axis. The spatial resolution differs between the x- and y-axes. This discrepancy likely arises from the fact that, while the scintillator spacing is uniform along the x-axis, it becomes slightly narrower along the y-axis due to sagging under the weight of scintillators. The resulting denser packing in the $y$-direction may therefore show the superior spatial resolution. After applying the corresponding position-dependent corrections, the re-reconstructed hit map (Fig.~\ref{fig:corrected_hitmap}) shows markedly improved agreement with the DWC reference. The “cross-shaped” structure visible in the hit map corresponds to the $5\times5$ arrangement of the CsI scintillator blocks. Although the beam is aimed at a single target block, its finite size and scattering cause partial illumination of the neighbouring blocks along the vertical and horizontal directions. This results in enhanced counts along the corresponding rows and columns, producing the observed cross-like pattern.

\section{Conclusion and Outlook}\label{sec:conculsion}
SAKURA has demonstrated that simple radiation imaging across various particle types can be achieved at a cost of approximately 1,000 USD. Since the introduction of Cosmic Watch in 2018—a single scintillator and SiPM-based educational detector—many institutions have adopted it, enabling students to easily conduct basic radiation measurements. However, radiation imaging has remained largely inaccessible in educational settings due to the need for multiple sensors corresponding to the number of pixels in an image.
\par
With SAKURA, even high school students can now perform simplified forms of SPECT (Single-Photon Emission Computed Tomography) or muography at home. In fact, all stages of this study—from detector development and construction to beam testing and data analysis—were carried out by high school students. By utilising Red Pitaya, an FPGA-based ADC board widely used in educational contexts, students can reconstruct measured radiation images without treating the processes of data acquisition and signal processing as a "black box", even if they have no prior experience in coding or data analysis. This makes imaging studies possible in classrooms or at home.

The next step is to expand opportunities for high school students to explore radiation imaging using this detector. This will be made possible through the educational outreach network of Accel Kitchen \cite{AccelKitchen}, which has supported more than 200 students by lending over 300 units of Cosmic Watch detectors for use in home-based experiments. These efforts have already led to a variety of student-led research outcome. Notably, several students have attempted pseudo-radiation imaging using a single Cosmic Watch, for instance, constructing a simple CT scanner or estimating the structure of school buildings via muography. With SAKURA, even more such inquiry-based learning experiences are expected to emerge.

\section*{Acknowledgements}
We gratefully acknowledge the support of the Beamline for Schools program at CERN, which provided the opportunity and resources to test our detector using the T10 beamline. We wish to extend our particular gratitude to the organisers, Sarah Z\"ochling and Markus Joos, for their support in coordinating the program. We also express our sincere gratitude to the Mitsubishi Mirai Foundation for their generous financial support.

\bibliographystyle{ptephy}
\bibliography{refs}

\begin{thebibliography}{10}

\bibitem{CosmicWatch}
S.N. Axani, K.~Frankiewicz, and J.M. Conrad,
\newblock The cosmicwatch desktop muon detector: a self-contained, pocket sized particle detector, Journal of Instrumentation, {\bf 13}(03), P03019 (Mar 2018), {https://doi.org/10.1088/1748-0221/13/03/P03019}.

\bibitem{Radiacode}
Radiacode Ltd.,
\newblock \url{https://www.radiacode.com} (2025),
\newblock Accessed: April 19, 2025.

\bibitem{appliedPhys_sakura}
{\begin{CJK*}{UTF8}{min}松下 千穂里、貫輪 美博、澤井 愛実、佐々木 柚榎、跡部 蒼、河野 理夏子、田中 香津生\end{CJK*}},
\newblock \begin{CJK*}{UTF8}{min}{CERN}の高校生実験コンテスト，{Beamline for Schools} 2024に参加して\end{CJK*} [{Participating in CERN's High School Physics Competition, Beamline for Schools 2024}], \begin{CJK*}{UTF8}{min}応用物理\end{CJK*}, {\bf 94}(5), 232--236 (2025), {https://doi.org/10.11470/oubutsu.94.5\_232}.

\bibitem{appliedPhys_tanaka}
{\begin{CJK*}{UTF8}{min}田中 香津生\end{CJK*}},
\newblock \begin{CJK*}{UTF8}{min}中高生が放射線探究を楽しむ世界を目指して [{Towards an Environment Where Middle and High School Students Engage in Radiation Research}]\end{CJK*}, \begin{CJK*}{UTF8}{min}応用物理\end{CJK*}, {\bf 94}(5), 249--253 (2025), {https://doi.org/10.11470/oubutsu.94.5\_249}.

\bibitem{PhysRev.75.590}
W.~C. Barber,
\newblock {East-West Asymmetry and Latitude Effect of Cosmic Rays at Altitudes up to 33,000 Feet}, Phys. Rev., {\bf 75}, 590--599 (Feb 1949), {https://doi.org/10.1103/PhysRev.75.590}.

\bibitem{PhysRev.93.590}
Charles~E. Miller, Joseph~E. Henderson, David~S. Potter, Jay Todd, Wayne~M. Sandstrom, Gerald~R. Garrison, William~R. Davis, and Francis~M. Charbonnier,
\newblock {The Zenith Angle Dependence of Cosmic-Ray Protons}, Phys. Rev., {\bf 93}, 590--595 (Feb 1954), {https://doi.org/10.1103/PhysRev.93.590}.

\bibitem{Morishima2017}
Kunihiro Morishima et~al.,
\newblock {Discovery of a big void in Khufu's Pyramid by observation of cosmic-ray muons}, Nature, {\bf 552}(7685), 386--390 (Dec 2017), {https://doi.org/10.1038/nature24647}.

\bibitem{Procureur2023}
S{\'e}bastien Procureur et~al.,
\newblock {Precise characterization of a corridor-shaped structure in Khufu's Pyramid by observation of cosmic-ray muons}, Nature Communications, {\bf 14}(1), 1144 (Mar 2023), {https://doi.org/10.1038/s41467-023-36351-0}.

\bibitem{Tanaka2019}
Hiroyuki K.~M. Tanaka,
\newblock Japanese volcanoes visualized with muography, Philosophical Transactions of the Royal Society A: Mathematical, Physical and Engineering Sciences, {\bf 377}(2137), 20180142 (2019), {https://doi.org/10.1098/rsta.2018.0142}.

\bibitem{LoPresti2020}
D.~Lo~Presti et~al.,
\newblock {Muographic monitoring of the volcano-tectonic evolution of Mount Etna}, Scientific Reports, {\bf 10}(1), 11351 (Jul 2020), {https://doi.org/10.1038/s41598-020-68435-y}.

\bibitem{Tanaka2010}
Hiroyuki K.~M. Tanaka, Tomohisa Uchida, Manobu Tanaka, Hiroshi Shinohara, and Hideaki Taira,
\newblock Development of a portable assembly-type cosmic-ray muon module for measuring the density structure of a column of magma, Earth, Planets and Space, {\bf 62}(2), 119--129 (Feb 2010), {https://doi.org/10.5047/eps.2009.06.003}.

\bibitem{ANGHEL201512}
V.~Anghel et~al.,
\newblock A plastic scintillator-based muon tomography system with an integrated muon spectrometer, Nuclear Instruments and Methods in Physics Research Section A: Accelerators, Spectrometers, Detectors and Associated Equipment, {\bf 798}, 12--23 (2015), {https://doi.org/https://doi.org/10.1016/j.nima.2015.06.054}.

\bibitem{Saracino2017}
G.~Saracino et~al.,
\newblock {Imaging of underground cavities with cosmic-ray muons from observations at Mt. Echia (Naples)}, Scientific Reports, {\bf 7}(1), 1181 (Apr 2017), {https://doi.org/10.1038/s41598-017-01277-3}.

\bibitem{PhysRevResearch.2.023017}
L.~F. Thompson, J.~P. Stowell, S.~J. Fargher, C.~A. Steer, K.~L. Loughney, E.~M. O'Sullivan, J.~G. Gluyas, S.~W. Blaney, and R.~J. Pidcock,
\newblock Muon tomography for railway tunnel imaging, Phys. Rev. Res., {\bf 2}, 023017 (Apr 2020), {https://doi.org/10.1103/PhysRevResearch.2.023017}.

\bibitem{10.1063/1.4922006}
J.~M. Durham, E.~Guardincerri, C.~L. Morris, J.~Bacon, J.~Fabritius, S.~Fellows, D.~Poulson, K.~Plaud-Ramos, and J.~Renshaw,
\newblock Tests of cosmic ray radiography for power industry applications, AIP Advances, {\bf 5}(6), 067111 (Jun 2015), {https://doi.org/10.1063/1.4922006}.

\bibitem{Guardincerri2017}
Elena Guardincerri, Charlotte Rowe, Emily Schultz-Fellenz, Mousumi Roy, Nicolas George, Christopher Morris, Jeffrey Bacon, Matthew Durham, Deborah Morley, Kenie Plaud-Ramos, Daniel Poulson, Diane Baker, Alain Bonneville, and Richard Kouzes,
\newblock {3D Cosmic Ray Muon Tomography from an Underground Tunnel}, Pure and Applied Geophysics, {\bf 174}(5), 2133--2141 (May 2017), {https://doi.org/10.1007/s00024-017-1526-x}.

\bibitem{cern_bl4s_winners_2024}
{CERN},
\newblock {Students from Estonia, Japan and the USA win 11th edition of Beamline for Schools} (June 2024),
\newblock Press Release.

\bibitem{jofscieggs-cherenkov}
Y.~Kuze, K.~Kubota, N.~Kobayashi, and K.S. Tanaka,
\newblock How best to detect cherenkov lights, Journal of Science EGGS, {\bf 7}(2410003), 1 (Jan 2024).

\bibitem{AccelKitchen}
{Accel Kitchen},
\newblock \url{https://accel-kitchen.com/} (2021),
\newblock Accessed: August 6, 2025.

\end{thebibliography}
%

\end{document}